\documentclass[prd,amsmath,amssymb,twocolumn]{revtex4-2}
\pdfoutput=1
\usepackage[utf8]{inputenc}
\usepackage{amsmath}
\usepackage{comment}
\usepackage{graphicx}
\usepackage{siunitx}
\usepackage{tikz}
\usepackage{hyperref}
\usetikzlibrary{positioning}
\usepackage{float}
\usepackage{subcaption}
\usepackage{appendix}

\begin{document}

\title{Bayesian Time Delay Interferometry}

\author{Jessica Page}
\affiliation{Space Science Department, University of Alabama in Huntsville, 320 Sparkman Drive, Huntsville, AL 35899, USA}
\author{Tyson B. Littenberg}
\affiliation{NASA Marshall Space Flight Center, Huntsville, AL 35812, USA}

\begin{abstract}
Laser frequency noise (LFN) is the dominant source of noise expected in the Laser Interferometer Space Antenna (LISA) mission, at $\sim$7 orders of magnitude greater than the typical signal expected from gravitational waves (GWs). Time-delay interferometry (TDI) suppresses LFN to an acceptable level by linearly combining measurements from individual spacecraft delayed by durations that correspond to their relative separations. Knowledge of the delay durations is crucial for TDI effectiveness. The work reported here extends upon previous studies using data-driven methods for inferring the delays during the post-processing of raw phasemeter data, also known as TDI ranging (TDIR). Our TDIR analysis uses Bayesian methods designed to ultimately be included in the LISA data model as part of a ``Global Fit'' analysis pipeline. Including TDIR as part of the Global Fit produces GW inferences which are marginalized over uncertainty in the spacecraft separations and allows for independent estimation of the spacecraft orbits. We demonstrate Markov Chain Monte Carlo (MCMC) inferences of the six time-independent delays required in the rigidly rotating approximation of the spacecraft configuration (TDI 1.5) using simulated data. The MCMC uses fractional delay interpolation (FDI) to digitally delay the raw phase meter data, and we study the sensitivity of the analysis to the filter length. Varying levels of complexity in the noise covariance matrix are also examined. Delay estimations are found to result in LFN suppression well below the level of secondary noises and constraints on the armlengths to $\mathcal{O}(30)\ {\rm cm}$ over the ${\sim}2.5\ {\rm Gm}$ baseline.
\end{abstract}

\maketitle

\section{Introduction}\label{introduction}

Gravitational waves (GW) are now an important contributor to our understanding of the universe and have already unlocked many developments since their momentous first detection in 2015. Space-based GW detection with the Laser Interferometer Space Antenna (LISA) opens the possibility of detecting the majority of expected GW sources which lie in the 0.1 mHz -- 0.1 Hz band \cite{LISA}. Targeting a launch in the 2030s, LISA will measure the GW induced optical pathlength change between freely falling test masses using laser beams exchanged between a rotating array of three separate spacecraft in heliocentric orbit with mean inter-spacecraft distances of 2.5 million km \cite{LISA}. These distances make it impossible to keep the ``arms'' of the interferometer equal, leaving the laser frequency fluctuations $\frac{\Delta \nu}{\nu}$ present in the data at a fractional frequency amplitude of $\SI{e-13}{Hz^{-1/2}}$ while typical GW signals are found near $\SI{e-21}{Hz^{-1/2}}$, burying signals by six to eight orders of magnitude. 

Time-delay interferometry (TDI) is a well-studied method to suppress laser frequency noise (LFN) sufficiently below the level necessary to achieve the mission science goals. TDI works by applying linear combinations of the interferometric data series shifted by the light travel time (i.e ``delay times'' or ``delays'' or ``armlengths'' throughout this paper) between spacecraft to digitally represent an equal-length interferometer, manifesting a data product with LFN common to all points of interference, effectively cancelling the overwhelming LFN and leaving the signal information intact \cite{TDI_original}. TDI requires knowledge of the delay times at each inter-spacecraft interference point which are functions of the separations and will vary on the order of $\sim$ 1 $\mathrm{m}/\mathrm{s}$. Delay time input to the TDI equations requires an accuracy estimated to range from $\sim$ 3--100 ns (or 1--30 m) depending on the sample rate, assumptions on the motion of the LISA constellation and interferometric measurement configuration to suppress LFN adequately \cite{TDI_requirements, Wang_et_al_2015}. A psuedo-random noise laser phase modulator will be encoded into the phasemeter measurements sent to Earth for use in extracting the spacecraft separations in the pre-data processing step, which also involves synchronization of the clocks on-board each spacecraft \cite{Wang_et_al_2015}. TDI is reviewed extensively in Ref. \cite{TDI_review} and recent developments include new TDI combinations using spacecraft velocities and accelerations that numerically solve for delays \cite{Muratore_et_al_2020}, a Doppler shift scaling operation on TDI combinations for implementation of frequency domain phase measurements \cite{Bayle_Staab_2021} and demonstration of the coupling between the on-board anti-aliasing filter and the time dependence of the delays in TDI performance \cite{flexing_filtering}.

However as demonstrated in \cite{TDIR}, it is possible and beneficial to suppress LFN in the post-processing of the raw data products, rather than relying on the pre-processed ranging information which could reduce
the amount of inter-spacecraft communication required, serve as redundancy for the ranging measurement, and provide an alternative method for noise mitigation \cite{FDI}. This method is known as Time-Delay Interferometric Ranging (TDIR) and has been shown to successfully suppress LFN by determining the delay parameters required for TDI by minimizing the noise power in the raw data. The average light travel time error was shown to range from $\sim$0.8--133 ns depending on the integration time and orbit model used \cite{TDIR}. Various recent studies have also demonstrated methods of data-driven LFN suppression. Principal component analysis in the frequency domain was implemented to produce LFN-free data products and independent delay estimation using raw phasemeter measurements as input to the likelihood function in the LISA data model in Ref.~\cite{Baghi_et_al_2021}. A related example of direct use of the interferometer measurements rather than transformed TDI variables in Ref.~\cite{TDI_infinity} generalizes the time dependence of the delays without truncating the expansion to first or second order.

A key component to TDIR, as demonstrated in Ref.~\cite{TDIR} and built upon in this work, is fractional delay interpolation (FDI), shown to work well for LISA requirements~\cite{FDI}. Data is telemetered to Earth at a sampling rate approximately $10^{7}$ times less than what is required for 100 ns accuracy of the delays. FDI reconstructs the signal at the intermediate time steps between samples that correspond to the delays by convolving the signal with a windowed-sinc filter. The filter length corresponds to the amount of data that must be removed from the interpolated segment since only the portion where the filter was completely immersed in the convolution is usable. Less data loss is better for LISA since interruptions in data acquisition are expected. Ref. \cite{FDI} found that at a sampling rate of 10 Hz using a LaGrange window with filter length $N = 15$ meets the interpolation error requirements to suppress the LFN below the secondary noises. We explore methods in section \ref{TDIR_MCMC} to reduce data loss due to FDI and show that a smaller estimate of the oversampling required for FDI is achievable in section \ref{filter}.

\begin{figure*}
\begin{subfigure}{.5\textwidth}
  \includegraphics[width=1.0\linewidth]{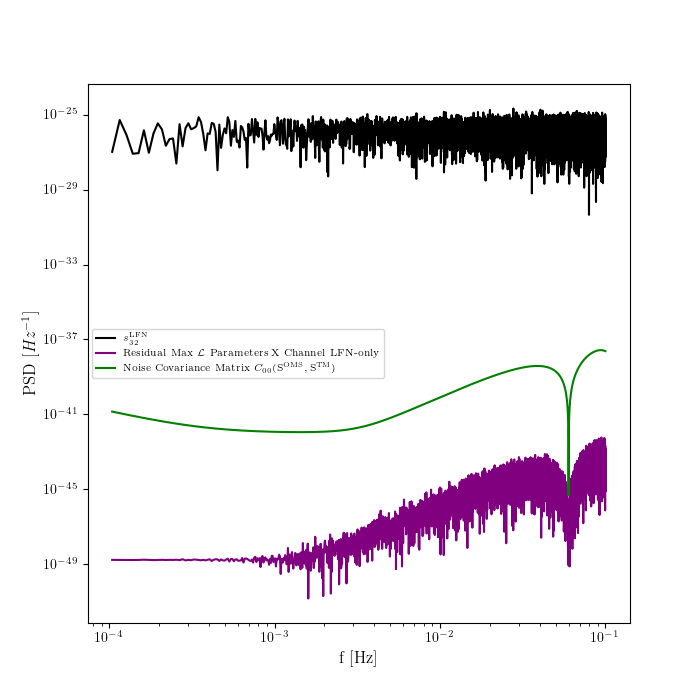}
  \caption{}
  \label{fig:residual_LFN_only}
\end{subfigure}%
\begin{subfigure}{.5\textwidth}
  \includegraphics[width=1.0\linewidth]{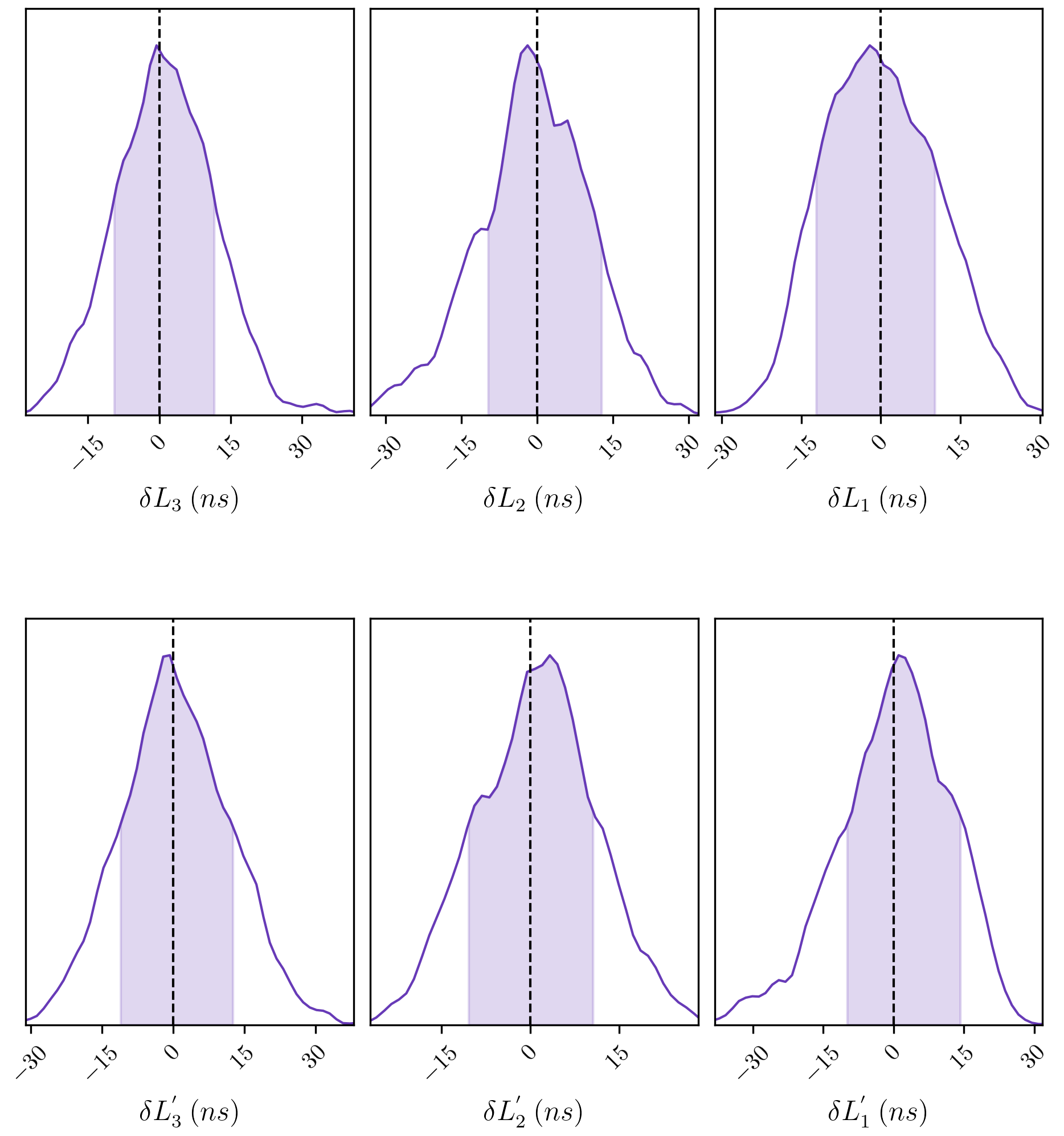}
  \caption{}
  \label{fig:one_D_dist_LFN_only}
\end{subfigure}
\caption{(a) TDI $X$ channel residual of the LFN using maximum likelihood parameters from the MCMC posterior (purple) compared to analytic optical metrology (OMS) and test mass (TM) noise PSD (green) which is an approximate representation of the GW sensitivity. The LFN component to the science measurement $\mathrm{s}_{32}$ (black) is also shown to demonstrate LFN input to spacecraft \#1 measurements before TDI suppression. (b) Marginalized posterior distribution functions of the six independently measured delay estimates in nanoseconds for data containing LFN only. The distributions are centered on the true value for the armlength.}
\label{fig:full_LFN_only}
\end{figure*}

The work presented here aims to extend upon the ideas of TDIR noise minimization using Bayesian probabilistic methods, namely a Markov Chain Monte Carlo (MCMC) algorithm to infer the delays from the raw phase measurements. The product is a package-able function that can be incorporated into the various GW search algorithms being developed for the LISA mission. Bayesian methods in gravitational wave data analysis have proved highly successful in the interpretation of LIGO/Virgo detections \cite{LALInference, BayesWave, BayesLine, Bilby}. A brief example of the algorithm's success in LFN suppression is shown in figure \ref{fig:full_LFN_only} and is further discussed in section \ref{LFN_suppresion}. The black curve is an example of raw science measurement data containing LFN that has been suppressed (purple) to well below the secondary noises (green) by the TDI $X$ channel using the maximum likelihood point estimates from the posterior distribution function sampled by the MCMC. The right hand panel of the figure shows the marginalized posterior distribution functions for the six inferred armlengths (measured separately on the outward and return directions between each spacecraft pair). 

Inferences of astrophysical signal properties are made using parameterized models of the LISA response to incident GWs, which include the detector armlengths in the modeling. This introduces the opportunity to include TDIR delay estimation into GW signal extraction models by treating the delays as additional parameters in these models and inferring their values by what optimally agrees with the data. This also allows for an independent estimation of the orbital motion of the spacecraft and a natural way to marginalize over uncertainty in the spacecraft separations. This approach to TDIR will also be used to measure the degree to which uncertainty in the ranging measurement can exist without disrupting the GW information. Second generation (TDI 2.0) LFN cancellation which accounts for at least the first order time dependence of the delays requires knowledge of  the relative spacecraft velocities which is often given by an analytic model for the spacecraft orbits. Another benefit to this data-driven ranging is the option to generalize to arbitrary orbital motion where the velocities are treated as additional parameters in the model and solved for using techniques that integrate along the path length being measured. This is essentially a technique that solves for the relative orbital motion of the LISA constellation. By exploring various configurations, we can improve efficiency by finding what best yields the correct parameters in the analysis. 

Before the recent split interferometry design for LISA \cite{otto_2012}, armlength accuracy of $\leq$ 30 ns was expected to be achievable under a 100 ns requirement  (\cite{TDIR},\cite{FDI}). We update the armlength accuracies attainable using the split interferometry design in section \ref{LFN_suppresion}. In the remainder of this paper, we will demonstrate the performance of the MCMC TDIR algorithm on simulated data using a rigid rotating approximation to the LISA orbits.
The paper is organized as follows: Section \ref{sec_2} describes how LFN enters in the LISA measurement system and provides the MCMC algorithm details. Section \ref{sec_3} covers the results and discussion of the performance testing on simulated data. We conclude and outlook to future use of the algorithm in section \ref{conclusion}.

\section{TDIR Implementation Using MCMC Sampling}\label{sec_2}

\subsection{LISA Measurement Description}\label{LISA_measurement}

The LISA measurement system is assumed here to operate under the \textit{split interferometry} design described in \cite{otto_2012} which combines three interferometric measurements to extract the GW signal. Each spacecraft contains two test masses and two optical benches that send and receive signals in both the clockwise and counter clock-wise directions of the rotating spacecraft array. Labeling conventions vary widely across TDI studies, but we follow that of \cite{reference_TDIR} and add the modifications necessary for split interferometry. The science measurement $s_{ij}$ containing the GW signal is the laser light received from the optical bench on distant spacecraft $i$ that is adjacent to the local optical bench $j$. The test mass measurement $\varepsilon_{ij}$ measures the motion of the test mass on spacecraft $j$ adjacent to spacecraft $i$, and the reference measurement $\tau_{ij}$ interferes the laser received on the optical bench adjacent to spacecraft $i$ on spacecraft $j$ with the other optical bench on spacecraft $j$.

\begin{figure}
    \includegraphics[scale=0.7]{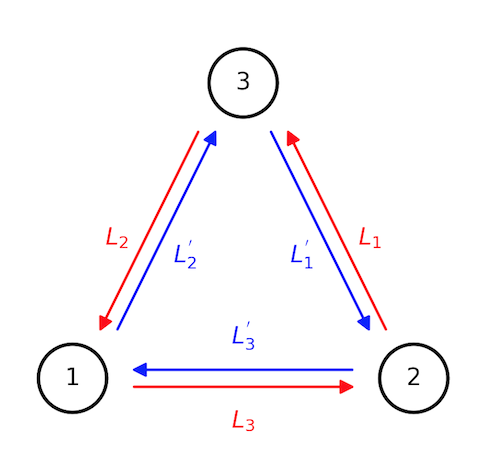}
    \caption{Schematic showing direction of primed and un-primed delays. Primed delays are clock-wise, un-primed delays are counter clock-wise.}
    \label{fig:diagram}
\end{figure}

Viewing the array face-on and following the diagram in Fig. \ref{fig:diagram}, spacecraft separations in the clock-wise direction ($L^{'}_{i}$) are indicated with a prime and counter clock-wise directions are un-primed ($L_{i}$). Because the array is rotating, the distance between two spacecraft will differ depending on the direction the laser light is traveling due to the Sagnac effect ($L_i \neq L^{'}_{i}$). Spacecraft separations are also time-dependent with $\dot{L_{i}}{\sim}$\SI{1}{m/s} but we ignore this for now and assume the array is in rigid rotation, therefore requiring the ``flex-free'' \cite{TDI_review} version of TDI (``TDI 1.5'' \cite{Baghi_et_al_2021}) only for this analysis. $p_{ij}$ is the laser frequency noise on the optical bench adjacent to spacecraft $i$ on spacecraft $j$. Laser light received in a science measurement $s_{ij}$ has laser noise $p_{ji,k}$ that has been delayed by the light travel time $L_{k}/c$ from spacecraft $i$ to $j$ combined with $p_{ij}$. The $_{,k}$ notation indicates the time delay operator $D_{k}$ that acts on a function $f(t)$ as $D_{k}f(t) = f(t-\frac{L_{k}}{c})$. Optical bench motion noise is delayed in the science measurements in the same form as laser noise and is assumed here to be suppressed along with LFN in the analysis, but is included as a contribution to the overall optical metrology system (OMS) noise. The remaining OMS noise (readout, shot, etc.) is assumed to be contained in the science measurement only and is denoted by $n^{\textrm{OMS}}_{ij}$. Test mass (TM) motion noise entering into $\varepsilon_{ij}$ measurements is combined in the term $n^{\textrm{TM}}_{ij}$. Ignoring the gravitational wave signal and focusing mainly on LFN suppression here, the six measurements on spacecraft 1 are as follows in (\ref{eq:measurments}). Cyclic permutation of both indices yields the remaining 12 measurements. 

\begin{align}
    s_{21} &= p_{12,3^{'}}-p_{21} + n^{\textrm{OMS}}_{21} \nonumber \\
    \tau_{21} &= p_{31}-p_{21} \nonumber \\
    \varepsilon_{21} &= p_{31}-p_{21} - 2 n^{\textrm{TM}}_{21} \nonumber \\
    s_{31} &= p_{13,2}-p_{31} + n^{\textrm{OMS}}_{31} \nonumber \\
    \tau_{31} &= p_{21}-p_{31} \nonumber \\ 
    \varepsilon_{31} &= p_{21}-p_{31} + 2 n^{\textrm{TM}}_{31} \label{eq:measurments}
\end{align}


\subsection{Data Model and MCMC Implementation}\label{TDIR_MCMC}
The goal of the MCMC is to estimate the delays applied to the first generation (rigidly rotating) Michaelson ``XYZ'' TDI combinations \cite{TDI_review} for effective LFN suppression. Cyclic permutation of indices in $X(t)$ in Eq. \ref{eq:TDI_equation} is applied for $Y(t)$ and $Z(t)$. The parameters estimated are all delay operators $\hat{k}$ that appear after a comma. 

\begin{widetext}
\begin{equation}\label{eq:TDI_equation}
\begin{split}
X(t)& = (s_{31}+s_{13,\hat{2}}) + (s_{21}+s_{12,\hat{3}^{'}})_{\hat{2}^{'}\hat{2}}  - (s_{21}+s_{12,\hat{3}^{'}}) - (s_{31}+s_{13,\hat{2}})_{\hat{3}\hat{3}^{'}}\\& +\frac{1}{2}((\tau_{21}-\tau_{31})_{\hat{2}^{'}\hat{2}\hat{3}\hat{3}^{'}} - (\tau_{21}-\tau_{31})_{\hat{3}\hat{3}^{'}} - (\tau_{21}-\tau_{31})_{\hat{2}^{'}\hat{2}} + (\tau_{21}-\tau_{31}))\\ &+ \frac{1}{2}((\varepsilon_{31}-\tau_{31})_{\hat{2}^{'}\hat{2}\hat{3}\hat{3}^{'}} + (\varepsilon_{31}-\tau_{31})_{\hat{3}\hat{3}^{'}} -  (\varepsilon_{31}-\tau_{31})_{\hat{2}^{'}\hat{2}} - (\varepsilon_{31}-\tau_{31}))\\ &-\frac{1}{2}((\varepsilon_{21}-\tau_{21})_{\hat{2}^{'}\hat{2}\hat{3}\hat{3}^{'}} - (\varepsilon_{21}-\tau_{21})_{\hat{3}\hat{3}^{'}} + (\varepsilon_{21}-\tau_{21})_{\hat{2}^{'}\hat{2}} - (\varepsilon_{21}-\tau_{21}))\\ &+ (\varepsilon_{12}-\tau_{12})_{\hat{3}^{'}} - (\varepsilon_{12}-\tau_{12})_{\hat{3}^{'}\hat{2}^{'}\hat{2}} - (\varepsilon_{13}-\tau_{13})_{\hat{2}} + (\varepsilon_{13}-\tau_{13})_{\hat{2}\hat{3}\hat{3}^{'}}
\end{split}
\end{equation}
\end{widetext}

Our TDIR analysis uses Gibbs sampling for the individual armlength parameters in a Metropolis-Hastings MCMC algorithm that utilizes Bayes' Theorem (\ref{eq:bayes}) to estimate the posterior distribution $p(\vec{\theta}|\vec{d})$ of the delay parameters $\vec{\theta} = \{\hat{L}_1,\hat{L}_2...\}$ given the data $\vec{d}$. FDI using a Lagrange-windowed sinc filter is applied at each iteration of the chain to precisely determine the time series delayed by the proposed armlength.

\begin{equation}\label{eq:bayes}
p(\vec{\theta}|\vec{d})\propto p(\vec{d}|\vec{\theta})p(\vec{\theta})
\end{equation}

We assume a uniform prior $p(\vec{\theta}) = \mathcal{U}[L_{\textrm{min}},L_{\textrm{max}}]$ where $L_{\textrm{min}}$ and $L_{\textrm{max}}$ are the approximate minimum and maximum armlengths in lightseconds over the course of a year in orbit. The data $\vec{d} = \{\tilde{X}(f),\tilde{Y}(f),\tilde{Z}(f)\}$ are the Fourier transforms of the TDI channels using the proposed delay parameters as input at each MCMC iteration. The Gibbs sampler draws each individual parameter separately from a Gaussian distribution centered on the last accepted value for a given parameter in units of time with a standard deviation that alternates between $10^5/c$, $10^3/c$, $10^2/c$ and $1/c$ seconds. 50000 steps were taken for all chains presented here except those in section \ref{LFN_suppresion} which used 100000 steps. $\mathcal{O}(10^{4})$ samples are taken on one CPU in around half the time of the duration of the data being analyzed. 

Equation (\ref{eq:bayes}) is implemented in logarithmic space, and the log likelihood function (\ref{eq:likelihood}) for the complex-valued random variables in $\vec{d}$, neglecting constant normalization factors, is 
\begin{equation}\label{eq:likelihood}
    \ln p(\vec{d}|\vec{\theta}) = \sum_{i=f_\textrm{min}}^{f_{\textrm{max}}} \left[-\ln(|\mathbf{C}|)_{i}-\left(\sum\limits_{j,k}^{X,Y,Z}\mathbf{d}_j^{\dagger}\mathbf{C}^{-1}_{jk}\mathbf{d}_k\right)_i\right],
\end{equation}
where $f_\textrm{min} = 10^{-4}\ {{\rm Hz}}$ and $f_\textrm{max}$ is restricted to 0.1 Hz instead of 1 Hz which is typically the assumed maximum of the LISA band. Interpolation error increases dramatically at frequencies near the 1 Hz maximum, so by restricting the maximum band of the likelihood function to 0.1 Hz (which is just below the transfer frequency $f_{*} = c/L$), we find in section \ref{filter} that better parameter accuracy is achieved using a lower sampling rate and significantly shorter filter length. Reducing the size of the filter length is desirable as it results in less data loss each time the signal is interrupted.

The noise covariance matrix $\mathbf{C}(f)$ describes the TM and OMS noises present in the $\mathbf{d}_j^{\dagger}$ and $\mathbf{d}_k$ terms (i.e. $\tilde{X}(f)$, $\tilde{Y}(f)$ and $\tilde{Z}(f)$ channels). All noises on each test mass and optical bench are independent and assumed to be stationary. We explore three versions of the covariance matrix at varying levels of complexity. The first is the set of noise orthogonal ``$AET$'' channels \cite{original_AET} which are combinations of the $XYZ$ channels that result in a GW-insensitive $T$ channel and, under the assumption of an equal arm configuration, results in a diagonal noise covariance matrix. We examined whether the equal-arm assumption in the noise covariance matrix had any effect on the arm length parameter estimation by deriving an unequal-arm XYZ noise covariance matrix  ($\mathbf{C}_{\mathrm{UNEQUAL}}$) to compare with results using the simplified equal-arm XYZ matrix ($\mathbf{C}_{\mathrm{EQUAL}}$) and the AET matrix ($\mathbf{C}_{\mathrm{AET}}$). Comparisons are described in section \ref{noise_matrix} and although the complexity of the noise covariance matrix is not expected to have a significant effect on LFN supression, we tested whether it would have a significant effect on armlength recovery since accurate armlengths could affect the GW parameter estimation. We provide the unequal XYZ matrix for the rigidly rotating (TDI 1.5) scenario in Eqs. (\ref{eq:unequal_1}) and (\ref{eq:unequal_2}) of the Appendix for potential need in the future. The diagonal elements of the equal-arm XYZ matrix are $\mathrm{C}^{\mathrm{EQUAL}}_{jj} = 16\sin^2{(2\pi f L)} S^{\mathrm{OMS}}_y+(8\sin^2{(4\pi f L)}+32\sin^2{(2\pi f L)}) S^{\mathrm{OMS}}_y$ and the off-diagonal elements are $\mathrm{C}^{\mathrm{EQUAL}}_{jk} = (4S^{\mathrm{Tm}}_y+S^{\mathrm{OMS}}_y) (-4 \sin{(2\pi f L)} \sin{(4\pi f L)})$. The AET matrix contains only the diagonal terms where $\mathrm{C}^{\mathrm{AET}}_{00} = \mathrm{C}^{\mathrm{AET}}_{11} = \mathrm{C}^{\mathrm{EQUAL}}_{jj} - \mathrm{C}^{\mathrm{EQUAL}}_{jk}$ and  $\mathrm{C}^{\mathrm{AET}}_{22} = \mathrm{C}^{\mathrm{EQUAL}}_{jj} + 2\mathrm{C}^{\mathrm{EQUAL}}_{jk}$. 

The first term of Eq. (\ref{eq:likelihood}) is a constant throughout all iterations of the MCMC for the equal-arm XYZ and AET implementations, but is updated for each proposed sample of the chain for the unequal-arm XYZ version since $\mathbf{C}$ is updated with each proposed armlength. The second term is the $\chi^2$ portion of the likelihood function which is used to examine the data generation method in section \ref{data_simulation} and the filter length tests in section \ref{filter}.

\section{Performance Testing}\label{sec_3}

\subsection{Data Simulation}\label{data_simulation}

To test the effectiveness of our TDIR algorithm, day-long segments of data sampled at 4 Hz were simulated. Laser frequency noise is white noise in the LISA band $10^{-4}-1$ Hz with fractional frequency power spectral density (PSD) $S^{\rm LFN}_{y} = 10^{-26}{\mathrm{Hz}^{-1}}$. The remaining test mass (TM) motion and optical metrology system (OMS) noise PSDs \cite{LISA} converted to fractional frequency are  

\begin{widetext}
\begin{align}
    S^{\textrm{TM}}_{y}(f) &= (2\pi f c)^{-2} (\num{3e-15})^2 \left(1+\left(\frac{\num{4e-4}}{f}\right)^2\right)  \left(1+\left(\frac{f}{\num{8e-3}}\right)^4\right)  \textrm{Hz}^{-1}\\
    S^{\textrm{OMS}}_{y}(f) &= \left(\frac{2 \pi f}{c}\right)^{2} (\num{1.5e-11})^2 \left(1+\left(\frac{\num{2e-3}}{f}\right)^4\right) \textrm{Hz}^{-1}.
\end{align}
\end{widetext}

\begin{figure*}
\begin{subfigure}{.5\textwidth}
  \includegraphics[width=1.0\linewidth]{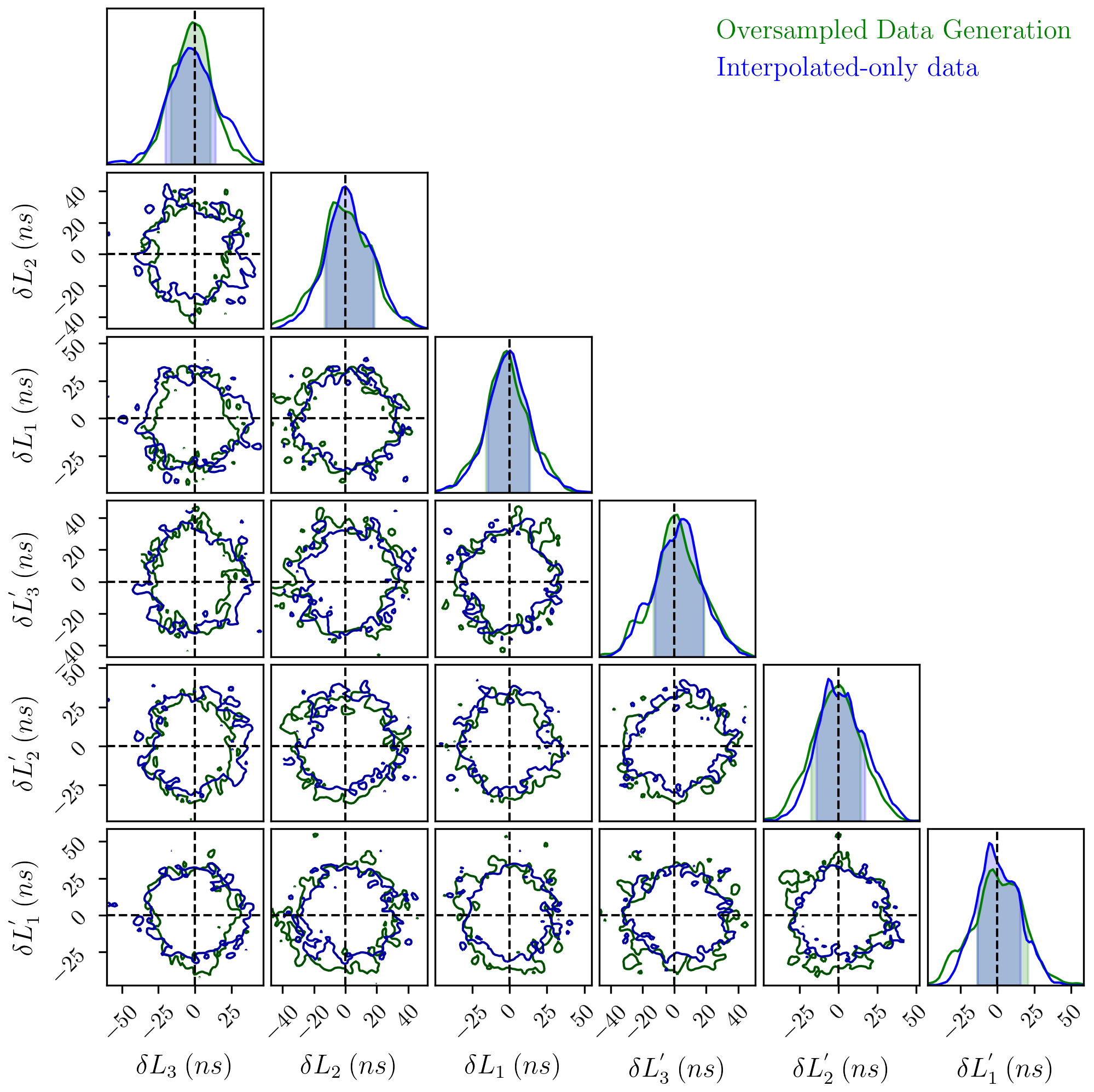}
  \caption{}
  \label{fig:data_gen_comparison}
\end{subfigure}%
\begin{subfigure}{.5\textwidth}
  \includegraphics[width=1.0\linewidth]{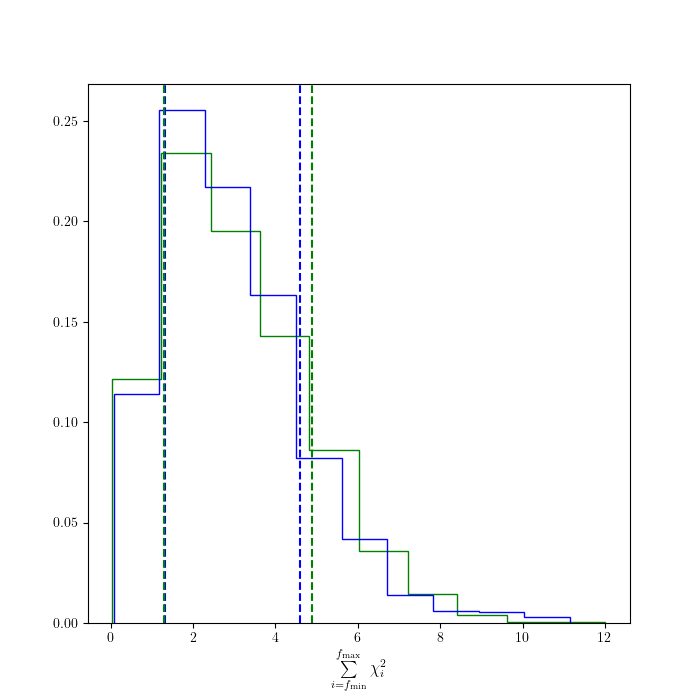}
  \caption{}
  \label{fig:chi_2_comparsion_data_generation}
\end{subfigure}
\caption{(a) Comparison of MCMC posteriors for the two data generation methods. Corner plots are arm-length posteriors subtracted by the true values used in the data generation. Shaded regions are 90\% credible interval. Comparing over-sampled plus interpolated data using an initial sampling rate of $\num{1e3}$ Hz and N=101 downsampled to 4 Hz (green) to interpolated-only data using the LaGrange filter with N=49 (blue). (b) Comparison of the $\chi^{2}_{i}$ sum over frequency bins distributions for the two data generation methods.}
\label{fig:data_generation}
\end{figure*}

Because we use FDI both for data simulation and estimating the delays, we first confirmed that the LFN suppression was not falsely successful due to cancellation of common interpolation error. 
The minimum filter length required for a given sampling rate as given in \cite{FDI} is determined by the maximum interpolation error in the LISA band which has a threshold of $10^{-8}$ for LISA specifications. We used the maximum error equation given in \cite{FDI} and calculated the minimum LaGrange filter length required for a sampling rate of 4 Hz and found N $\geq$ 49. Data generated using the LaGrange window with N=49 was compared to data beginning with a 1 kHz sampling rate using a LaGrange window with N=101 and then down-sampled to 4 Hz. Both oversampling by a factor of 500 and using a filter length higher than necessary is assumed to be a sufficient comparison to data generated at the sampling rate of the analysis. Arm-length posteriors are compared in Fig. \ref{fig:data_gen_comparison}. The $\chi^{2}$ term of likelihood function for the two data sets containing laser noises only (Fig. \ref{fig:chi_2_comparsion_data_generation}) are also included to show that the likelihood values are unchanged as well, demonstrating that the use of the same interpolation scheme for data simulation and data analysis does not mislead our conclusions.

\subsection{Interpolation Filter Performance} \label{filter}
\begin{figure*}[htp]
    \includegraphics[width=\textwidth]{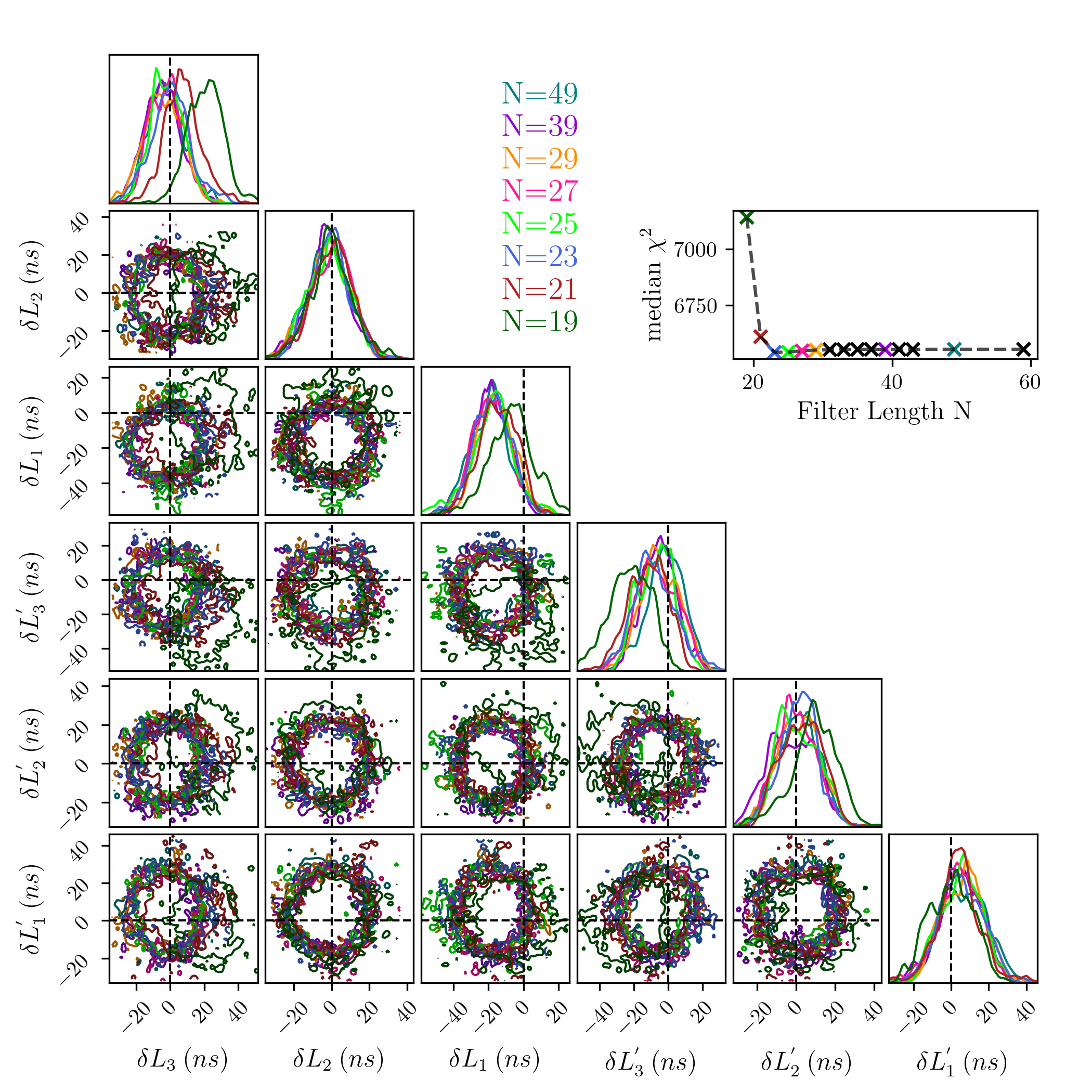}
    \caption{Delay posteriors at varying LaGrange window filter lengths with their corresponding median $\chi^2$ values. Posterior estimates are given as the difference from truth values used in the data simulation in nanoseconds. Data was sampled at 4 Hz using LaGrange filter length N=49. By restricting the frequency band in the likelihood to below 0.1 Hz, all filter lengths tested above N=21 achieve similar delay estimates. N=29 marks the beginning of the asymptotic portion of the filter length vs $\chi^2$ distributions.}
    \label{fig:filter_length_tests}
\end{figure*}

We examined which part of the band gave the maximum interpolation error for a 4 Hz sampling rate and found that, for all LaGrange-windowed filter lengths tested between 3 and 201, the maximum interpolation error was always due to frequency bins above 0.1 Hz and was usually near 1 Hz. We found that restricting the maximum frequency bin summed over in the likelihood function to 0.1 Hz resulted in a few orders of magnitude greater accuracy in the armlength estimation, allowing us to test whether a shorter filter length is possible when the part of the band with the highest interpolation error is excluded. While the simulated data used to test the MCMC armlength recovery used the theoretical best case N=49 for the LaGrange-windowed filter under a sampling rate of 4 Hz, the MCMC algorithm ran over varying filter lengths to determine the minimum filter length possible. We tested the LaGrange filter over $N=59$, 49 and the remaining odd integers decreasing from 43 to 19. The armlength posteriors in units of deviation from the expected values used in the simulation in nanoseconds and their corresponding $\sum_i \chi^{2}_i$ posteriors are shown in Fig. \ref{fig:filter_length_tests}. 

Some filter lengths were excluded from the corner plot in the figure for clarity, but we found that all filter lengths above N=21 recovered the same delay distributions. While $N=23$ is the minimum filter length required before parameter accuracy degradation, the $\chi^2$ behavior between $N=23$ and 27 does not match the higher filter lengths tested so we adopted the filter length of $N=29$ where the flat asymptotic trend in the median $\chi^2$ begins. This results in 7.25 s of data loss compared to 12.25 s using $N=49$. Data losses from interpolation will accumulate over the multi-year mission lifetime when interruptions cause gaps of missing samples. Minimizing data losses due to filtering is especially important for the long duration LISA sources that will remain in band for the entirety of the mission.

The 0.1 Hz cut-off in the likelihood function also allows for a lower sampling rate since the oversampling factor of 5 requirement for FDI \cite{FDI} was estimated for mitigation of high frequency aliasing. Data bandlimited to 1 Hz saw sufficient LFN suppression using an oversampling factor of 2, and a sampling rate of 3 Hz was also found to be comparable. The cubed Blackman filter mentioned in \cite{FDI} can use a slightly shorter filter length than the LaGrange filter at 4 Hz, but the Lagrange filter resulted in a smoother and smaller amplitude LFN residual. 

\subsection{Parameterized Noise Covariance Matrix} \label{noise_matrix}

\begin{figure*}
\begin{subfigure}{.5\textwidth}
  \includegraphics[width=1.0\linewidth]{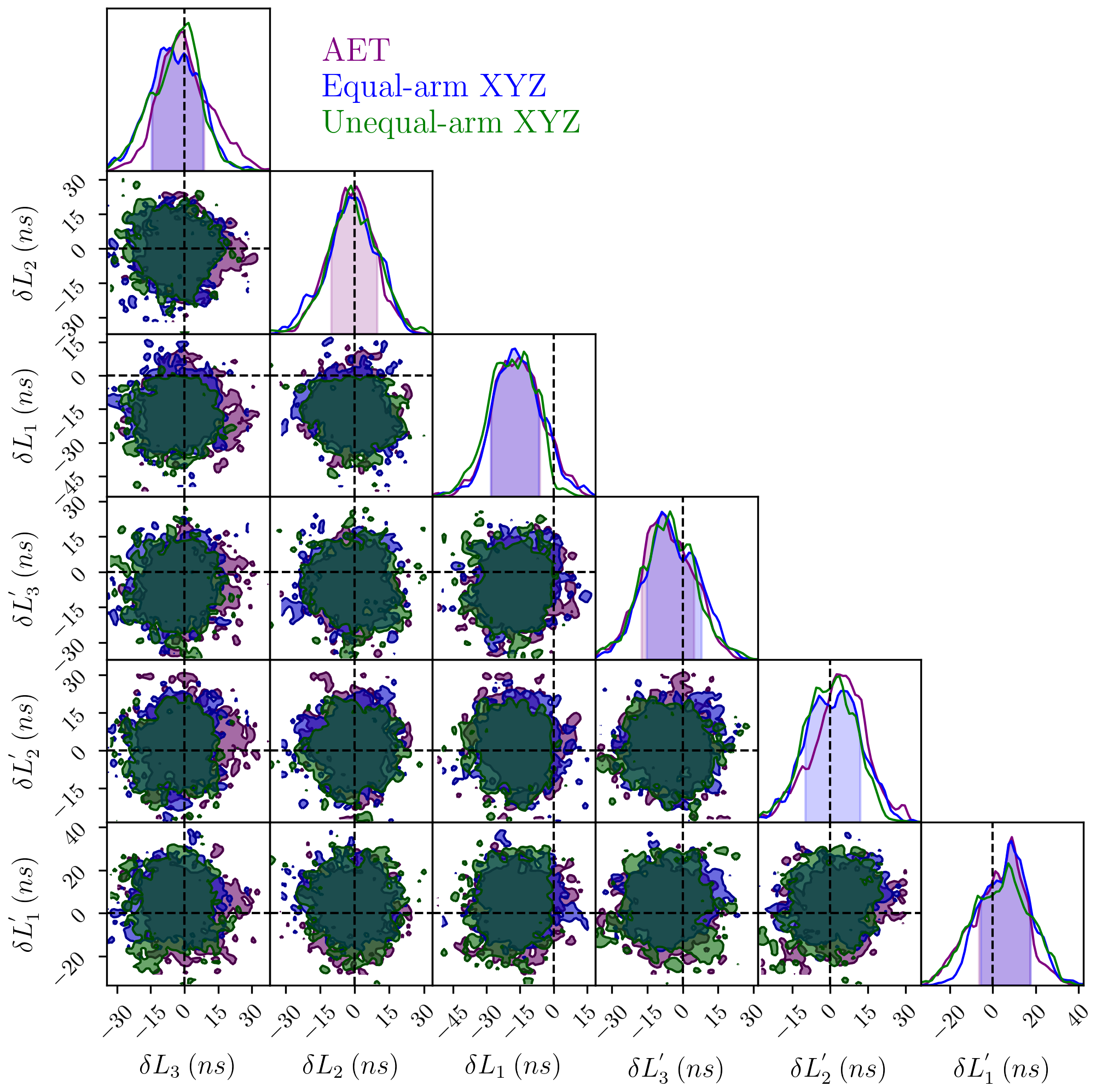}
  \caption{}
  \label{fig:noise_cov_matrix_corner_plot}
\end{subfigure}%
\begin{subfigure}{.5\textwidth}
  \includegraphics[width=1.0\linewidth]{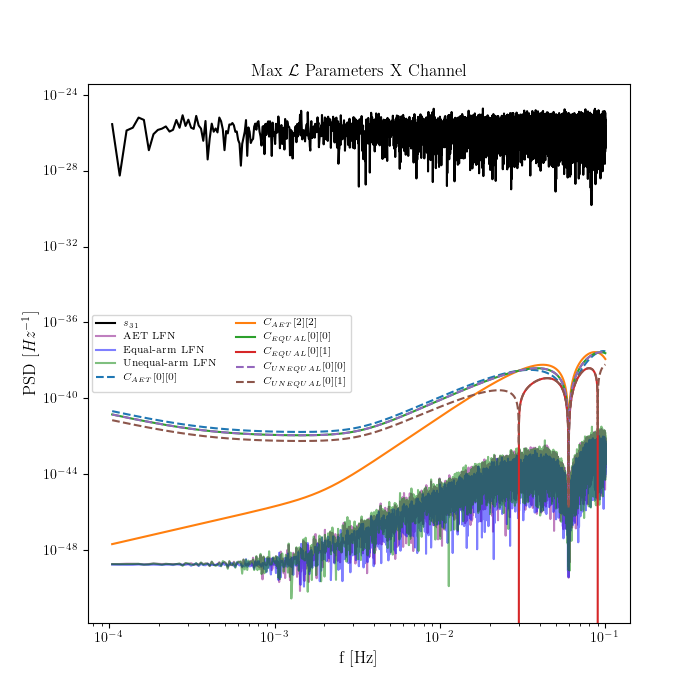}
  \caption{}
  \label{fig:noise_cov_matrix_residuals}
\end{subfigure}
\caption{(a) Posterior delay time estimates using the AET, equal and unequal arm versions of the noise covariance matrix. Estimates are given as the difference from truth values used in the data simulation in nanoseconds. No significant differences in estimates were found. (b) Posterior delay time point estimates from the maximum of the likelihood distribution as input to the $X$ channel. $s_{31}$ (black) is  shown as an example of the data. The resulting residual of the LFN portion of the data is shown for the AET (purple), equal (blue) and unequal arm (green) versions of the noise covariance matrix. Placement of the (0,0) and (0,1) elements of each matrix are shown which are representative of the entire matrix.}
\label{fig:full_noise_cov_comparison}
\end{figure*}

Using an interpolation filter length of $N=29$, we examined whether including the delay parameters in the noise covariance matrix had any affect on delay estimation or LFN suppression. Figure \ref{fig:noise_cov_matrix_corner_plot} compares the delay posteriors in deviations from truth values in nanoseconds for the AET, equal-arm XYZ and the full unequal-arm XYZ. No significant differences were found in either parameter estimation accuracy (Table \ref{tab:noise_compare}) or LFN suppression, (Fig. \ref{fig:noise_cov_matrix_residuals}) so the LFN suppression does not appear to be sensitive to changes in armlength in the noise covariance matrix. However we will revisit this comparison in the future when we use the algorithm alongside other analyses with data containing high signal-to-noise GW signals. Waveform models are dependent on armlength parameters, so we can study whether GW parameter estimation could benefit from the unequal-arm noise covariance matrix.

\renewcommand{\arraystretch}{2}
\begin{table*}
    \begin{tabular}{ccccccc}
        \hline
		Covariance Matrix & $\delta L_{3} \: (ns)$ & $\delta L_{2} \: (ns)$ & $\delta L_{1} \: (ns)$ & $\delta L^{'}_{3} \: (ns)$ & $\delta L^{'}_{2} \: (ns)$ & $\delta L^{'}_{1} \: (ns)$ \\ 
		\hline
		AET & $-0.653^{+22.21}_{-16.9}$ & $-1.02^{+15.39}_{-17.77}$ & $-15.95^{+2.51}_{-33.13}$ & $-7.03^{+11.08}_{-24.55}$ & $4.29^{+20.54}_{-16.26}$ & $4.23^{+20.71}_{-18.72}$ \\
		Equal-arm XYZ & $-4.57^{+13.28}_{-23.04}$ & $-1.26^{+15}_{-21.58}$ & $-16.2^{+1.89}_{-32.68}$ & $-5.38^{+14.09}_{-22.76}$ & $1.45^{+19.05}_{-15.9}$ & $7^{+24.4}_{-11.43}$ \\
		Unequal-arm XYZ & $-2.84^{+12.33}_{-22.35}$  & $-1.01^{+16.94}_{-18.55}$  & $-18.31^{-3.38}_{-36.11}$ & $-5.93^{+12.34}_{-24.27}$ & $0.26^{+15.85}_{-17.15}$ & $4.49^{+24.55}_{-18.7}$ \\
		\hline
    \end{tabular}

    \caption{Median and 90\% credible interval delay estimates for the AET, equal and unequal arm versions of the noise covariance matrix. Estimates are given as the difference from truth values used in the data simulation in nanoseconds. Numbers correspond to Fig. \ref{fig:noise_cov_matrix_corner_plot}.}
    \label{tab:noise_compare}
\end{table*}

\subsection{Including Secondary Noises} \label{LFN_suppresion}

\begin{figure*}
\begin{subfigure}{.5\textwidth}
  \includegraphics[width=1.0\linewidth]{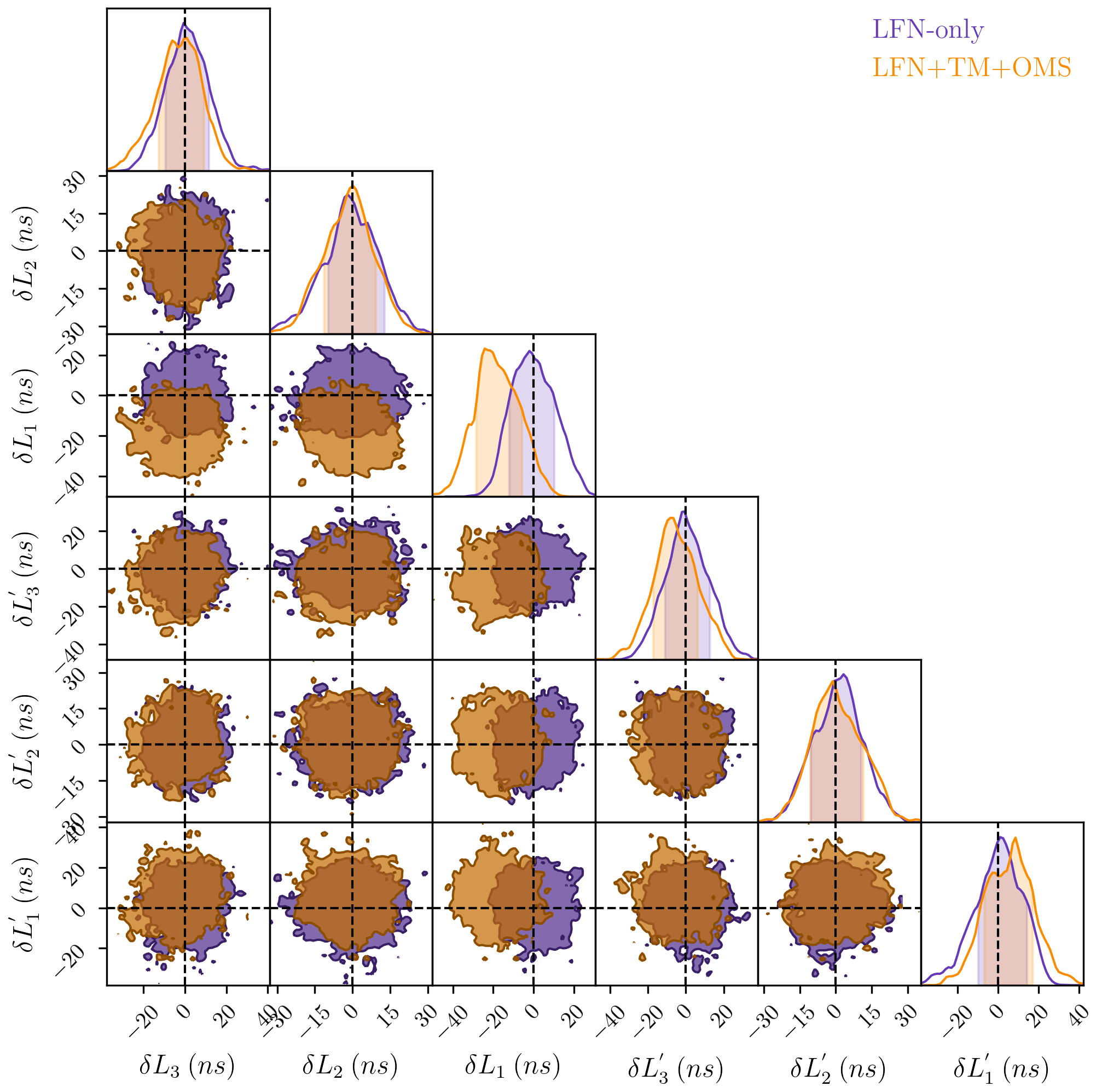}
  \caption{}
  \label{fig:LFN_only_vs_secondary}
\end{subfigure}%
\begin{subfigure}{.5\textwidth}
  \includegraphics[width=1.0\linewidth]{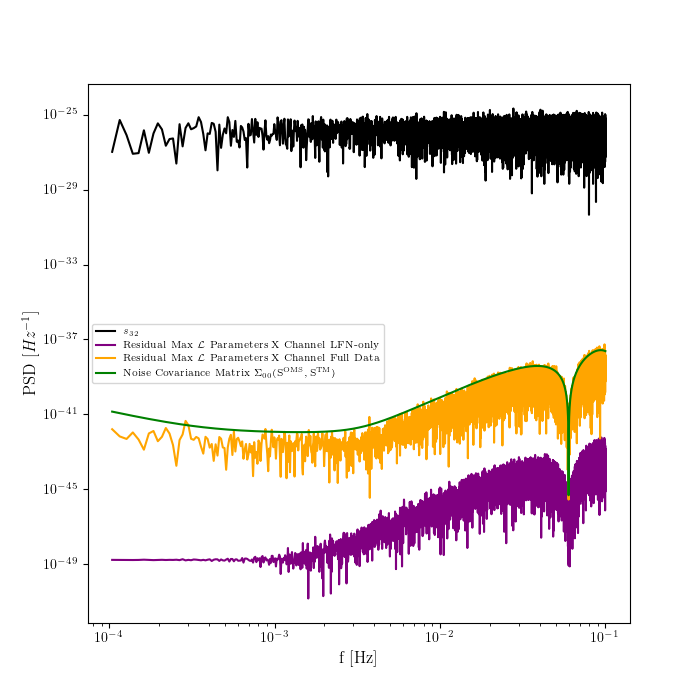}
  \caption{}
  \label{fig:LFN_vs_secondary_residual_BOTH}
\end{subfigure}
\caption{(a) Comparison of armlength posteriors using data that included TM and OMS noises with LFN (orange) vs data containing LFN only (purple). Estimates are given by sample values subtracted by the truth values that were used as input in the data simulation in nanoseconds. (b) $X$ channel residual of full data containing LFN+TM+OMS noise using maximum likelihood parameters from MCMC posterior (orange) compared to analytic OMS and TM noise PSD (green). $X$ channel residual of LFN data using maximum likelihood parameters from MCMC posterior (purple). The science measurement $\mathrm{s}_{32}$ (black) is also shown to demonstrate LFN input to spacecraft \#1 measurements before TDI suppression.}
\label{fig:full_noise_cov_comparison}
\end{figure*}

Concentrating on posteriors using a LaGrange filter length of $N=29$ under the equal-arm XYZ noise covariance matrix assumption, we take a closer look at the accuracy achieved in estimating the armlengths that are used as input to TDI to form the residual shown in Fig. \ref{fig:LFN_vs_secondary_residual_BOTH}. Our previous results did not include a realization for the secondary noises, and so we expect the minimum $\chi^2$ to be $0$ and, due to our flat priors and small correlations between parameters, expect the marginalized posteriors to peak at the true value for the armlengths. Now comparing posteriors from data that included secondary noises versus data containing LFN only, we see the size of the statistical error is comparable as expected (Fig. \ref{fig:LFN_only_vs_secondary}). Posteriors of data containing LFN only are centered on the input delays and data with the secondary noises are usually centered and are within ${\sim}1\sigma$. A more thorough analysis of the statistical robustness of our sampling algorithm is needed, but the sampler passes these qualitative tests which are sufficient at this stage of development.

Table \ref{table:model_params} summarizes the posteriors in figure \ref{fig:LFN_only_vs_secondary}, and we find that the least accurate parameter for data with secondary noises has a median accuracy of 18.51 ns or 5.5 m, and the most accurate parameter has a median accuracy of 0.07 ns or 2.1 cm. Most parameters with secondary noises were accurate to $\mathcal{O}$(1) ns or $\sim$ 30 cm. The deviation from truth appearing in the $\delta L_{1}$ posterior is due to the particular secondary noise realization in the data. This was confirmed by running the MCMC over multiple instances of simulated secondary noises where we found statistical fluctuations were random and not specific to one parameter. Using the set of parameters that resulted in the maximum likelihood of the chain as input to the TDI $X$ channel, the resulting residual is at the expected sensitivity for GW detection (Fig. \ref{fig:LFN_vs_secondary_residual_BOTH}). The LFN component to data that enters the $X$ channel is suppressed well below the secondary noises.

\begin{table*}
    \begin{tabular}{ccccccc}
        \hline
		Model & $\delta L_{3} \: (ns)$ & $\delta L_{2} \: (ns)$ & $\delta L_{1} \: (ns)$ & $\delta L^{'}_{3} \: (ns)$ & $\delta L^{'}_{2} \: (ns)$ & $\delta L^{'}_{1} \: (ns)$ \\ 
		\hline
		Secondary Noises Included & $-2.52^{+14.04}_{-22.96}$ & $-1.25^{+14.6}_{-18.4}$ & $-18.51^{+0.71}_{-35.52}$ & $-5.72^{+15.27}_{-24.37}$ & $-0.07^{+19.11}_{-15.85}$ & $6.15^{+25.07}_{-13.7}$ \\ 
		LFN-only & $-1.1^{+18.5}_{-16.1}$ & $-0.6^{+16.7}_{-19.7}$ & $-0.4^{+18.3}_{-16.1}$ & $0.3^{+19.8}_{-18.1}$ & $1.3^{+17.4}_{-15.9}$ & $0.8^{+18.4}_{-20.8}$ \\ 
		\hline
    \end{tabular}
    \caption{Median and 90\% credible interval delay accuracy for data that included the OMS and TM noises vs LFN-only data. Estimates are given as the difference from truth values used in the data simulation in nanoseconds.}
    \label{table:model_params}

\end{table*}

\section{Conclusion and Future Work}\label{conclusion}

We demonstrate Bayesian estimation of inter-spacecraft distances for use in post-processed TDI for the rigidly rotating LISA scenario using a Markov Chain Monte Carlo algorithm with fractional delay interpolation for the nanosecond precision delays. This data driven approach allows for flexibility in various key areas of LISA data analysis, and is shown to sample posteriors in less time than the duration of data being analyzed. The restricted bandwidth included in the likelihood function allows for a lower telemetered sample rate, a shorter FDI filter length requiring less data loss and higher accuracy in armlength parameter estimation to $\mathcal{O}(30)\ {\rm cm}$. The noise covariance matrix complexity does not affect accuracy or LFN suppression in GW-free data. 

The demonstrations in this paper used ideal stationary Gaussian noise simulations, and thus did not consider the impact of imperfections in the data. Two anticipated data imperfections for LISA are gaps and noise transients  or ``glitches''. The TDIR algorithm is designed to process short segments of data, adiabatically reconstructing the time-varying delays after analysis of several segments. In principal this approach is robust to gaps in the data, as the analysis can be performed between gaps with additional continuity constraints on the inferred delays from one segment to the next. The drawback to this approach is the loss of data at the beginning and end of each short segment due to the filtering, the impact of which will depend on the specific frequency and duration of data interruptions.

Glitches will appear in the data ``seen'' by the TDIR algorithm as excess broad-band noise, effectively raising the noise ``floor'' below which the LFN must be suppressed to minimize the likelihood.  If left unmitigated, glitches will primarily increase the uncertainty in the inferred delays. While the size of the affect glitches will have on TDIR should be studied, we ultimately envision the ranging inferences to be part of a global noise fit which includes a model for glitches similar to what has been used in the analysis of LIGO-Virgo data~\cite{BayesWave}.

Near-term future work on the TDIR algorithm will extend to time-dependent delays (TDI 2.0) for an analysis of the benefits of independent spacecraft orbit reconstruction and incorporate the ranging model into GW analysis algorithms such as Ref.~\cite{GBMCMC} to study the interplay between simultaneous recovery of armlengths and GW parameters.

\section{Acknowledgements}

We thank Q. Baghi, J. Baker, K. Lackeos, J. Slutsky and J. I. Thorpe for their helpful insights throughout the development of this work. The project was supported by the NASA LISA Study Office and NASA grant NNH15ZDA001N-APRA.

\appendix
\setcounter{secnumdepth}{0}
\section{Appendix: Unequal-Arm Noise Covariance Matrix}\label{cov_matrix_appendix}

The diagonal elements of $\mathbf{C}$ are $\mathrm{C}_{jj} = <|X_{j}(f)|^2>$ and off-diagonal elements are $\mathrm{C}_{jk} = <\overline{X_{j}(f)}X_{k}(f)>$ where $\mathbf{X} = \{X,Y,Z\}$. The TM and OMS components of the diagonal elements are listed separately for simplicity and then combined to form each matrix element ($\mathrm{C}_{jk} = \mathrm{C}^{\mathrm{PM}}_{jk} + \mathrm{C}^{\mathrm{OMS}}_{jk}$.) The matrix is Hermitian so only the first three off-diagonal elements are listed.

\begin{widetext}
\begin{align}\tag{A1}\label{eq:unequal_1}
\begin{split}
\mathrm{C}^{\mathrm{OMS}}_{00} &= 4 S^{\mathrm{OMS}}_{y}(f)\left(2-\cos{2\pi f(L_3+L^{'}_3)}-\cos{2\pi f (L_2+L^{'}_2)}\right)\\
\mathrm{C}^{\mathrm{OMS}}_{11} &= 4 S^{\mathrm{OMS}}_{y}(f)\left(2-\cos{2\pi f(L_1+L^{'}_1)}-\cos{2\pi f (L_3+L^{'}_3)}\right)\\
\mathrm{C}^{\mathrm{OMS}}_{22} &= 4 S^{\mathrm{OMS}}_{y}(f)\left(2-\cos{2\pi f(L_2+L^{'}_2)}-\cos{2\pi f (L_1+L^{'}_1)}\right)\\
\mathrm{C}^{\mathrm{TM}}_{00} &= -4 S^{\mathrm{TM}}_{y}(f)\left(-6+2\cos{2\pi f(L_2+L^{'}_2)} + \cos{2\pi f(L_2+L^{'}_2-L_3-L^{'}_3)} \right. \\& \left. +2\cos{2\pi f(L_3+L^{'}_3)}+ \cos{2\pi f(L_2+L^{'}_2+L_3+L^{'}_3)} \right)\\
\mathrm{C}^{\mathrm{TM}}_{11} &= -4 S^{\mathrm{TM}}_{y}(f)\left(-6+2\cos{2\pi f(L_1+L^{'}_1)} + \cos{2\pi f(L_1+L^{'}_1-L_3-L^{'}_3)} \right. \\& \left. +2\cos{2\pi f(L_3+L^{'}_3)}+ \cos{2\pi f(L_1+L^{'}_1+L_3+L^{'}_3)} \right)\\
\mathrm{C}^{\mathrm{TM}}_{22} &= -4 S^{\mathrm{TM}}_{y}(f)\left(-6+2\cos{2\pi f(L_1+L^{'}_1)} + \cos{2\pi f(L_1+L^{'}_1-L_2-L^{'}_2)} \right. \\& \left. +2\cos{2\pi f(L_2+L^{'}_2)}+ \cos{2\pi f(L_1+L^{'}_1+L_2+L^{'}_2)} \right)\\
\end{split}
\end{align}
\begin{align}\tag{A2}\label{eq:unequal_2}
\begin{split}
\mathrm{C}_{01} &= \left(4 S^{\mathrm{TM}}_{y}(f)+S^{\mathrm{OMS}}_{y}(f)\right) e^{-2 i \pi f(L_2+L^{'}_2+L^{'}_3)} \left(-1+e^{2 i \pi f(L_1+L^{'}_1)}\right)\left(-1+e^{2 i \pi f(L_2+L^{'}_2)}\right)\left(1+e^{2 i \pi f(L_3+L^{'}_3)}\right)\\
\mathrm{C}_{02} &= \left(4 S^{\mathrm{TM}}_{y}(f)+S^{\mathrm{OMS}}_{y}(f)\right) e^{-2 i \pi f(L_2+L_3+L^{'}_3)} \left(-1+e^{2 i \pi f(L_1+L^{'}_1)}\right)\left(1+e^{2 i \pi f(L_2+L^{'}_2)}\right)\left(-1+e^{2 i \pi f(L_3+L^{'}_3)}\right)\\
\mathrm{C}_{12} &= \left(4 S^{\mathrm{TM}}_{y}(f)+S^{\mathrm{OMS}}_{y}(f)\right) e^{-2 i \pi f(L^{'}_1+L_3+L^{'}_3)} \left(1+e^{2 i \pi f(L_1+L^{'}_1)}\right)\left(-1+e^{2 i \pi f(L_2+L^{'}_2)}\right)\left(-1+e^{2 i \pi f(L_3+L^{'}_3)}\right)
\end{split}
\end{align}
\end{widetext}

\bibliography{main}

\end{document}